\PassOptionsToPackage{usenames,dvipsnames,svgnames,table}{xcolor}
\documentclass[twoside,leqno,twocolumn]{article} % siam
\AtBeginDocument{%
  \providecommand\BibTeX{{%
    \normalfont B\kern-0.5em{\scshape i\kern-0.25em b}\kern-0.8em\TeX}}}

\usepackage{ltexpprt}

\usepackage[letterpaper,margin=2cm]{geometry} % siam

\usepackage{graphics} % siam
\usepackage{tikz}
\usepackage{tikz}
\usepackage{graphicx}
\usepackage{bm}

\usepackage{booktabs} % For formal tables
\usepackage{color}

\usepackage{caption}
\usepackage[normalem]{ulem}
\usepackage{tabularx}
\usepackage{algorithm}
\usepackage[noend]{algpseudocode}
\usepackage[caption=false]{subfig}
\usepackage{amsmath,amssymb}

\usepackage{mathtools}
\graphicspath{{./figures/}}

\algblock{ParFor}{EndParFor}
\algnewcommand\algorithmicparfor{\textbf{parallel for}}
\algnewcommand\algorithmicpardo{\textbf{do}}
\algnewcommand\algorithmicendparfor{}
\algrenewtext{ParFor}[1]{\algorithmicparfor\ #1\ \algorithmicpardo}
\algrenewtext{EndParFor}{\algorithmicendparfor}

\usepackage{color}
\usepackage{array}
\newcolumntype{L}[1]{>{\raggedright\let\newline\\\arraybackslash\hspace{0pt}}m{#1}}
\newcolumntype{C}[1]{>{\centering\let\newline\\\arraybackslash\hspace{0pt}}m{#1}}
\newcolumntype{R}[1]{>{\raggedleft\let\newline\\\arraybackslash\hspace{0pt}}m{#1}}
\usepackage{listings}

\makeatletter
\ifthenelse{\equal{\ALG@noend}{t}}%
  {\algtext*{EndOn}}
  {}%
\makeatother

\begin{document}

\title{Scalable Hash Table for NUMA Systems}

\author{Alok Tripathy\thanks{University of California, Berkeley. Work done while at NVIDIA.}
\and Oded Green\thanks{NVIDIA, Georgia Institute of Technology}}

% \author{Alok Tripathy}
% \authornotemark[1]
% \email{alokt@berkeley.edu}
% \affiliation{University of California, Berkeley}
% 
% \author{Oded Green}
% \email{ogreen@nvidia.com}
% \affiliation{NVIDIA, Georgia Institute of Technology }
% 
% \bstctlcite{IEEEexample:BSTcontrol}
% 
% \authornote{Authors worked on this project while at NVIDIA.}
% \affiliation{\institution{Georgia Institute of Technology}}
% \email{ogreen@gatech.edu, atripathy8@gatech.edu}

\maketitle

\begin{abstract}
Hash tables are used in a plethora of applications, including database operations, DNA sequencing, string searching, and many more.
As such, there are many parallelized hash tables targeting multicore, distributed, and accelerator-based systems.
We present in this work a multi-GPU hash table implementation that can process keys at a throughput comparable to that of distributed hash tables.
Distributed CPU hash tables have received significantly more attention than GPU-based hash tables. 
We show that a single node with multiple GPUs offers roughly the same performance as a $500-1,000$-core CPU-based cluster.
Our algorithm's key component is our use of multiple sparse-graph data structures and binning techniques to build the hash table.
As has been shown individually, these components can be written with massive parallelism that is amenable to GPU acceleration.
Since we focus on an individual node, we also leverage communication primitives that are typically prohibitive in distributed environments.
We show that our new multi-GPU algorithm shares many of the same features of the single GPU algorithm---thus we have efficient collision management capabilities and can deal with a large number of duplicates. 
We evaluate our algorithm on two multi-GPU compute nodes:  1) an NVIDIA DGX2 server with 16 GPUs and 2) an IBM Power 9 Processor with 6 NVIDIA GPUs.
With $32$-bit keys, our implementation processes $8B$ keys per second, comparable to some $500 - 1,000$-core CPU-based clusters and $4\times$ faster than prior single-GPU implementations. 
% All of this work is open-source on Github: \todo{link}.

\setlength{\textfloatsep}{0pt}
\end{abstract}

%
%  Use this command to print the description
%
%\printccsdesc

% \keywords{ACM proceedings; \LaTeX; text tagging}

\section{Introduction}
\label{sec:intro}
% Loo https://www.osti.gov/servlets/purl/1141691

With applications including $k$-mer counting in biology \cite{pan2018optimizing}, sparse matrix multiplication\cite{anh2016balanced}, graph triangle counting \cite{pandey2019h}, and inner-join operations\cite{balkesen2013main,blanas2011design,kim2009sort,KimSortvsHash}, hash tables are among the most widely used data structures.
Historically, hash tables have been deployed on an extensive set of architectures, from shared-memory with many-cores \cite{CagriSortvsHash,maier2016concurrent}, to massively multi-threaded systems \cite{goodman2010hashing}, GPU accelerated \cite{harris2007cudpp, khorasani2015stadium, ashkiani2016gpu, junger2018warpdrive, greenHashgraph2019, alcantara2009real}, and up to large scale systems distributed \cite{barthels2017distributed,pan2018optimizing,gilray2019distributed}.
As accelerator-based computing grows, in particular GPU accelerators,  the need for a single-node multiple-accelerator hash table continues to grow. 
For brevity and simplicity, we will use the term GPU in this paper in the context of our algorithm and use the term accelerator for only the broader-case where a GPU is not applicable. 

Multi-GPU systems have advanced to the point they have as much memory as a many shared-memory systems with the benefit of the GPU's higher throughput.
CUDA's support of unified memory means that having multiple GPUs within a given compute node increases memory size. 
In the case of an NVIDIA DGX-2 system, with 16 NVIDIA V100 GPUs, each with 32GB of memory, an individual node has 512GB of GPU memory. 
The IBM POWER9 AC922 system with 6 NVIDIA V100 GPUs can have up to 96GB of GPU memory. 
These GPU-based systems have an equal amount of memory found in many single-node CPU systems and have computational resources that distributed systems do not have. 
Specifically, the DGX-2 has NVIDIA's NVSwitch interconnect, and the IBM Power 9 has NVLink 2nd Generation. 
These fast interconnects with the addition of unified memory turn these systems into high-speed shared memory systems (despite a resemblance to a distributed system).
% The unified memory also enables using single-node multi-GPU systems to solve larger problems than can be solved on a single device. 
In this paper, we introduce a scalable hash table for single-node multi-GPU systems with performance comparable to that of $500-1,000$-core distributed CPU systems.

While GPU-based implementations are gaining interest and often outperform shared-memory algorithms, GPU implementations are bottlenecked by memory.
NVIDIA V100 GPUs, for instance, have 32 GB of high bandwidth memory (HBM2). 
An NVIDIA RTX8000 has 48GB of GDDR6 memory. 
Because of this limitation, multi-GPU implementations are necessary for GPU-implementations to process as much data as shared-memory implementations.
Single-node multi-GPU hash tables have received some attention lately, such as cuDPP\cite{harris2007cudpp}, StadiumHash \cite{khorasani2015stadium}, SlabHash  \cite{ashkiani2016gpu}, and WarpDrive \cite{junger2018warpdrive}.
However, out of all of these, only WarpDrive is multi-GPU.
Also, none of those mentioned above, leverage the wealth of work in sparse-graph data structures or binning techniques.
Our work is the first to use both sparse-graph structures and sophisticated binning for a multi-GPU hash table, and achieves higher throughput as a result.

Overall, in this paper, our contributions are the following.
\begin{itemize}
    \item We introduce a single-node multi-GPU static hash table implementation with performance comparable to that of small distributed CPU-based clusters. Specifically, our implementation can process $2^{33}$ keys at a rate of $8B$ keys / second on a DGX-2 system. We accomplish this by extending an existing single-GPU hash table.
    \item We show that our new algorithm shares many of the key features that the single-GPU algorithm has. This includes a highly efficient collision management technique that copes with a large number of duplicates without a loss in performance.
    \item We introduce a multi-GPU static hash table implementation that can process $16\times$ more keys than existing single-GPU hash tables. To this end, we leverage work in sparse-graph data structures and binning techniques not previously seen in distributed hash table work, to the best of our knowledge.
\end{itemize}

\setlength{\textfloatsep}{0pt}

\section{Background}
\label{sec:background}
% Hash tables are found in a wide array of applications requiring efficient lookups.  Specifically, given an input array $A$ of length $n$ and  a value of interest $x$, a hash table can determine whether the value $x$ is part of the input array $A$ in $O(1)$ operations. In contrast, finding if $x$ exists in $A$ assuming that $A$ is not sorted requires $O(n)$ operations and $O(\log (n))$ random memory operations if $A$ is sorted. 
% In many real-world applications, the input data is not sorted.
% Therefore, even if the data is sorted, a hash-table can prove to be more computationally efficient for lookups and queries. 

A hash table maps every element in the input into an integer index using a \textbf{hash} function. These values are placed in the hash-table using a specific set of rules. These rules determine how the hash table behaves and are the primary differentiating factor between hash tables. For example, the hash table rules help determine how to deal with collisions-- the case where the input has multiple instances of the same value.
These rules are also part of the querying as they determine the correct search pattern within the hash table. 

% \paragraph{\bf Hash Table Terminology}
Hash functions map elements (that can be integers, floating points, strings, or complex data-types) to integers. The hash function is  typically defined as follows, given an element $e \in A$ the hash of $e$ is:
\begin{align*}
    hash(e) \in \{0,1,2,3,... \ldots, n / C\}
\end{align*}
where $n$ is the number of input elements, and $C$ is known as the \textbf{load factor}. $n/C$ is the hash-range of the input. The underlying assumption of a good hash function is that the probability function of the output follows a good uniform distribution. In simple terms, given two different inputs $a$ and $b$, the likelihood of {\it hash(a) = hash(b)} is very small.
$ C $ (a real number) is typically determined by the user with a trade-off between reducing the number of collisions at the cost of increasing the memory footprint. For example, $C=0.5$ increases the hash range to be $2\cdot n$ and lowers chances of collision. In contrast, $C=0.9$ results in a smaller hash range and an increased chance that two input hash values will \textbf{collide}.
%Large values of $C$ require a large backing array, adding a large memory footprint.
%However, small values of $ C $ increase the chance of two elements mapping to the same hash value, known as a \textbf{collision}.

% Over the last four decades, significant research focused on designing hash functions, including Murmur \cite{appleby2008murmurhash} \todo{More hash functions}. 

In the remainder of the paper, we use the following terminology. A key, $e$ refers to an element that is pre-hash. A key can be part of the original input or part of the query set.
A hashed-value refers to a key that has been hashed and will be denoted by {\it hash(e)}. Some hash tables store only the key inside the table while other tables can store a key-value pair. The ``value'' an extra piece of data associated with the key; for example,  it can be the index in the original input array. This is common for ``join'' operations that require the index. 
% The key difference between a hash-table that stores keys versus a hash-table that stores key-value pairs is the table's memory footprint. 

\paragraph{\bf Collision Management}
Hash tables have different collision management techniques--- with trade-offs from programmability, storage overhead, and computational complexity. In this subsection, we cover several of these in more detail.
Many hash tables are an extension of two broad types of hash tables: separate chaining and open-addressing. 

\begin{figure*}[t!]
    \centering
    \subfloat[Separate chaining]{\label{fig:sep-chain}\includegraphics[scale=0.2]{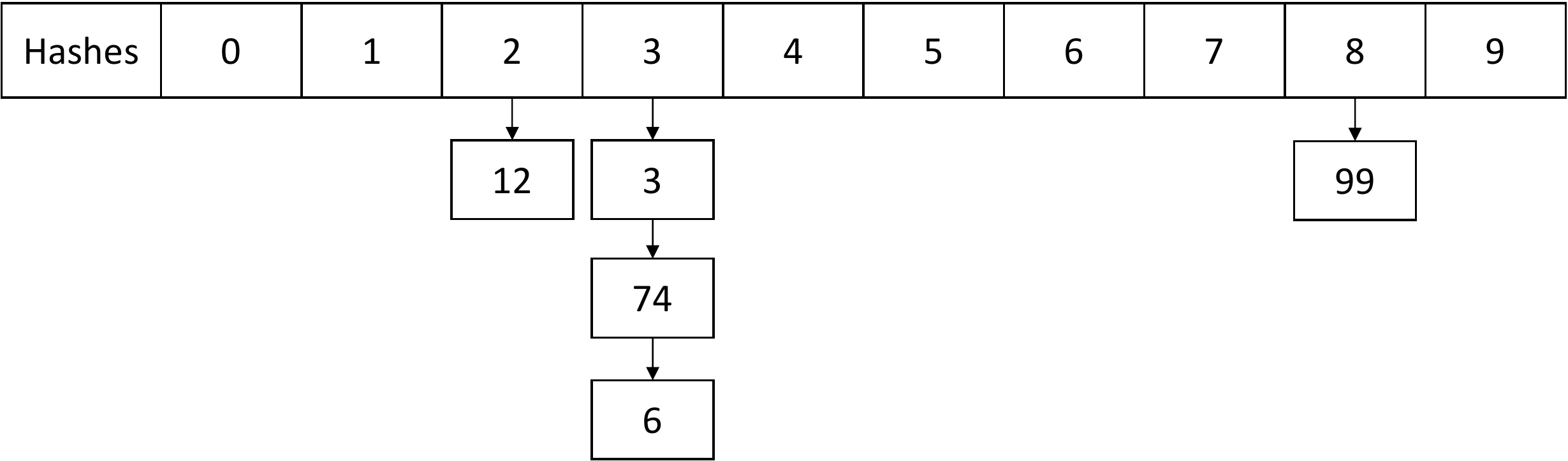}}
    % \qquad
    \hspace{3cm}
    \subfloat[Open addressing]{\label{fig:open-addr}\includegraphics[scale=0.2]{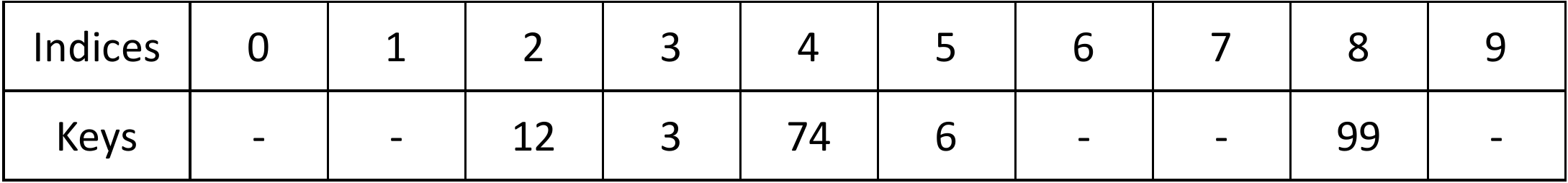}}
    \caption{Example hash tables built with keys $12, 3, 74, 6, 99$ using either separate chaining or open addressing collision management.}
    \label{fig:collision-strats}
\end{figure*}

\paragraph{Separate Chaining} Hash-tables based on the separate chaining collision approach, maintain a chain of keys hashed to a given hash value. Implementations typically use an array of linked lists such that each hash value has its own linked list (which are inherently sequential and limit scalability).
%Given a good hash function and assuming that the hash-range is roughly the same size as the input array, the size of the average linked list is small.
Figure \ref{fig:sep-chain} depicts this with the long linked list at index $3$. 
% Looking up any key that hashes to $3$ requires iterating over the whole linked list, which is inherently sequential and inefficient./
% The use of linked lists significantly reduces the performance of these tables.  Specifically, the lack of locality in the linked list limits parallel scalability as only one element can be accessed at a time. 
The lack of locality also means that memory accesses are random. Also, it doubles the memory footprint as we need to store both the keys and the next pointer in the list. 
% SlabHash by Ashkiani \emph{et. al.} \cite{ashkiani2016gpu} uses a linked-list of buckets such that each node in the linked list stores multiple elements-this improves scalability, improves locality, and cuts back on the next pointer's memory jump at the cost of having empty buckets.
%n contrast, the buckets can potentially be partially empty. Therefore, the performance of SlabHash is very dependent on the input.

\paragraph{Open Addressing} Many high-performance hash tables are open addressing based. 
One of the biggest attractions of implementing an open-addressed hash table is its simplicity and portability across architectures.
The entire table is stored within a single memory allocation. Specifically, the hash-range is also the hash table size as each hash value can store a single key. If the desired hash value is already in use, then the following entries in the hash table are scanned until an empty entry is found. 
% For open-addressing based hash-tables, it is very desirable to have a large hash-range as this reduces the number of collisions to the same hash-value and reduces the scan length in case of a collision. 
Open-addressing based methods fall short when the number of duplicates is high or when a large number of keys get hashed to the same hash value or vicinity. The cost of adding duplicate hash values grows at a quadratic rate with the number of duplicates. 
%For instance, if a value appears $d$ times in the input, it takes $O(d^2)$ scans to find $d$ empty entries in the hash-table. This gets compounded when multiple values get hashed to the same vicinity in the hash-table. By default, open-addressing hash-tables need additional logic in their implementations to deal with value deletions in the hash-table.
Figure \ref{fig:open-addr} depicts this, as inserting any key that hashes to $2$ requires traversing the keys with hash values $2$ and $3$.

\paragraph{Other Hash Tables}
Other collision management techniques include $2$-choice hashing \cite{richaTwohash2001}, and cuckooo hashing \cite{paghCuckoo2004}, and HashGraph \cite{greenHashgraph2019}.

\paragraph{\bf Static vs. Dynamic Hash Tables}
On top of collision management techniques, hash tables can also be categorized as \textit{static} or \textit{dynamic}.
In static hash tables, the entire set of input keys is specified upfront.
After building the hash table with these input keys, users can look up the values in this set, but they cannot add any additional keys or remove existing ones.
Dynamic hash tables contrast this by permitting insertions and deletions after building the hash table.
In this work, we focus on static hash tables, though our methods can be extended to the dynamic case.

% \paragraph{\bf Key vs. Key-Value Hash Tables}
% Hash tables can also be differentiated based on whether they store keys or both keys and associated values.
% For the former, hash tables only store keys (such as integers), and looking up a query key determines whether the query exists in the hash table.
% Key hash tables are also known as \textit{hash sets} in some contexts.
% In the latter, hash tables store an associated value for each key in the table.
% Lookups not only return whether a query key exists, but also return the key's value if it exists.
% While key-value hash tables are a superset of key hash tables, frequently only key hash tables are necessary, avoiding the overhead of having the store values.
% Our implementation focuses on key hash tables, though it can be extended for key-value hash tables as well.

\paragraph{\bf Parallel Hash Tables}
The need for fast and scalable hash tables in a plethora of applications has brought with it an abundance of parallel hash tables designed for multi-threaded, massively multi-threaded, and distributed systems. Performance for these hash tables is typically given with a throughput metric such as the number of keys per second.
Initially, many had worked on multicore CPU implementations of hash tables \cite{CagriSortvsHash,maier2016concurrent, goodman2010hashing}. 
These implementations were designed for systems with a few hundred threads at most. These hash tables tend not to scale to larger thread counts due to the use of a large amount of memory per thread or dependent on having large private caches.
% Changkyu \emph{et al.} \cite{kim2009sort} show a way to reduce the number of cache misses by maintaining numerous partitions per thread. Each thread is responsible for data reorganization using efficient cache accesses. Then the partition sized hash-table, one per threads, is built in a cache efficient manner. In a final step, the partition sized hash tables are merged to build the final output.  While this method is effective for a small number of threads, it is ineffective for hundreds to thousands of threads and is effectively a non-starter for tens of thousands of threads due to large memory overheads.

For instance, Maier \emph{et. al.} \cite{maier2016concurrent} propose the ``Folklore'' CPU-based hash table, which can build a table with roughly $300$ million keys per second.
Balkesen \emph{et. al.} \cite{CagriSortvsHash} also developed a CPU-based hash table, one that can build a table at a rate of $450$ million keys per second.
Goodman \emph{et. al.} \cite{goodman2010hashing} introduce CPU-based hashing strategies for the Cray XMT, which build at a table at roughly $250$ million keys per second.
All of these tables, however, do not provide the performance that distributed or accelerator-based implementation offers. 
Barthels et. al. \cite{barthels2017distributed} introduce two distributed inner-join algorithm, with the fastest running at roughly $48$ billion keys per second on $4096$ cores.

Hash-tables implementation for the GPU are very different than their CPU counterparts. Specifically, modern-day GPUs requires some 40k threads to keep the system fully utilized. This large thread count requires taking a very different approach as it is not possible to maintain a large amount of algorithmic state per thread. Another differentiating feature between CPUs and GPUs is that GPUs offer very efficient atomic operations.% The low overhead of the atomic instructions is an essential factor in the performance of the GPU based hash-tables.

For example, GPU based hash tables can build at a rate of over 1 billion keys per second on similar generation architectures. This includes the following hash tables:
cuDPP\cite{harris2007cudpp}, StadiumHash \cite{khorasani2015stadium},
SlabHash\  \cite{ashkiani2016gpu}, WarpDrive \cite{junger2018warpdrive},
HashGraph  \cite{greenHashgraph2019}. Moreover, while all these hash tables are faster than CPU based hash tables, they each take a very different approach in building the hash table.
%  usually the key differences  is in the collision management techniques used by these algorithms.
cuDPP uses Cuckoo hashing \cite{pagh2004cuckoo} to manage collisions. In the case of a collision, Cuckoo hashing will move the older value to a different place in the table (after rehashing) and store the new value in its place. 
% Thus, Cuckoo is a hybrid of open-addressing and separate chaining where the chains are created with numerous hash functions instead of being in a single, consecutive spot in the memory.
Cuckoo is very useful when the input follows a uniform distribution, and there are few collisions. StadiumHash is also Cuckoo based and targets large tables that do not fit inside the physical memory of the system.
%  Similar to cuckoo hashing, cuDPP has some performance benefits but suffers with too many duplicate keys.
WarpDrive is another recent GPU-based hash table that has both single GPU and multi-GPU implementations. The multi-GPU implementation used a multi-split to ensure good scalability. WarpDrive, unfortunately, does not permit duplicate keys in the input.

HashGraph by Green is a static hash table that uniquely treats hash tables as bipartite graphs \cite{greenHashgraph2019} where the two vertex sets are the  keys and their corresponding hash values. 
With this problem formulation, Green shows an approach for building the hash table in a cache friendly manner.
In contrast to open-addressing based tables, HashGraph avoids the extra work of finding an empty spot in the case of a collision.
Altogether, HashGraph outperforms all the aforementioned GPU-based hash tables on a single GPU.
Lastly, in contrast to many hash tables, HashGraph can cope with a huge number of collisions without losing performance -- we too build off this feature an ensure that our multi-GPU algorithm can deal with a large number of duplicates.

% The reason behind this lies with HashGraph's backing sparse-graph data structure.
% HashGraph uses a \textit{Compressed Sparse Row (CSR)} structure, a common sparse-graph data structure.
% At a high-level, CSR is effectively an adjacency list implemented with arrays in place of linked lists.
% For example, HashGraph can easily support deletions of values from the hash-table will the others easily support insertions. HashGraph can support insertions by using a dynamic graph data structure. The implementation in \cite{greenHashgraph2019} focuses on static inputs.

\setlength{\textfloatsep}{0pt}

% \section{Single-Accelerator HashGraph}
% \label{sec:single-gpu}
% \input{symbols}
% \setlength{\textfloatsep}{0pt}
% \input{single-gpu}
% \setlength{\textfloatsep}{0pt}

\section{Multi-GPU HashGraph}
\label{sec:multi-gpu}
\begin{figure*}[t!]
    \centering
    \subfloat[Bipartite graph representation.]{\label{fig:bip-graph}\includegraphics[scale=0.2]{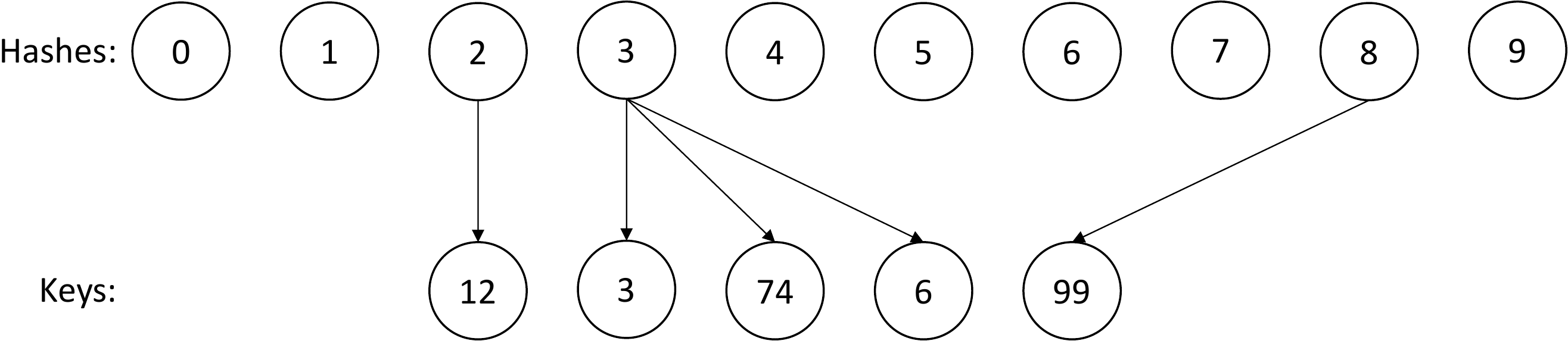}}
    % \qquad
    \hspace{3cm}
    \subfloat[CSR representation of graph.]{\label{fig:csr-repr}\includegraphics[scale=0.2]{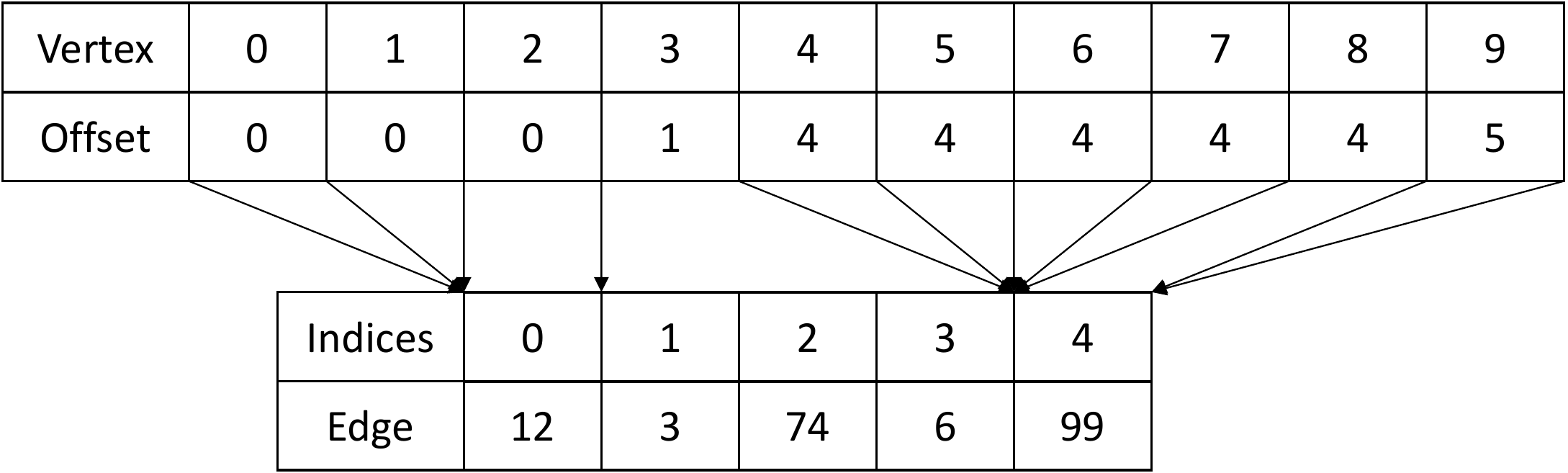}}
    \caption{Bipartite graph of hash values with their keys and the graph's corresponding CSR representation.}
    \label{fig:graph-csr}
\end{figure*}

In this section, we introduce our new multi-GPU algorithm. 
Our new algorithm is based on the HashGraph formulation, as discussed by Green \cite{greenHashgraph2019}. 
To fully understand our new multi-GPU algorithm, we cover the single-GPU version of HashGraph in the following subsections \ref{sec:single_gpu_build}.
After defining the terminology used by HashGraph, we will extend the single-GPU algorithm to multi-GPU in subsections. 
Our introduction of the single-GPU HashGraph is brief and covers the relevant details. 
We refer the reader to Green \cite{greenHashgraph2019} for additional details.

\subsection{Single-GPU HashGraph}
Green \cite{greenHashgraph2019} shows a new process for building a hash table by using a sparse-graph data structure, namely compressed sparse row (CSR). The HashGraph formulation shows that building a hash table is similar to storing it in a sparse bipartite graph. In Figure \ref{fig:graph-csr}, we show an example bipartite graph representation of keys and their hash values, along with the graph's CSR representation. 
% We discuss CSR in detail later in this section.

In the following subsections, we explain how to build a single-GPU HashGraph, yet we prelude by stating that HashGraph's uses an efficient collision management methodology. Rather than placing the element into the hash table in a single pass over the input data. HashGraph uses a sequence of parallel for loops to count the number of instances of hashed inputs into each hash value. With the number of instances for each hash value, a varying size memory-bin is set aside for each hash value. This process of determining the bin sizes uses a prefix sum operation similar to the offset calculation required by a CSR representation. After the allocation process, it is possible to efficiently place elements into the hash table using an additional pass over the input. The amortized cost for adding each item into the hash table is $O(1)$ regardless of the number of duplicates. HashGraph offers excellent performance when the number of copies in the hash table is high.  

% There are two different algorithms of HashGraph given by Green \cite{greenHashgraph2019}: 1) a naive algorithm that offers the intuition as to why the formulation is correct and 2) a highly efficient algorithm that uses caching and binning to reduce the number of random memory accesses. The second algorithm is significantly faster than the first for a single device. 
% In our multi-GPU HashGraph algorithm, we extend the formulation of the second single-GPU Hash graph algorithm. We chose this algorithm as the binning strategy offers a good starting point for scaling our hash table to multi-GPUs. We use a similar binning strategy, though, the purpose of our binning is different than the one used by Green \cite{greenHashgraph2019}.

Green \cite{greenHashgraph2019} shows two algorithms for building HashGraph: 1) a naive algorithm that offers intuition and 2) a highly efficient algorithm that uses caching and binning to reduce the number of random memory accesses. 
% The second algorithm is significantly faster than the first for a single device. 
In our multi-GPU HashGraph algorithm, we extend the formulation of the second single-GPU Hash graph algorithm. We chose this algorithm as the binning strategy offers a good starting point for scaling our hash table to multi-GPUs. We use a similar binning strategy, though, the purpose of our binning is different than the one used by Green \cite{greenHashgraph2019}.

\paragraph{Compressed-Sparse Row}
CSR is a well-known data structure used to represent sparse graphs and matrices. For the sake of completeness, we offer the following brief description of CSR. A CSR data structure typically consists of two arrays.
The lengths of these arrays are respectively $n$ and $m$ (or $ |V| $ and $ |E| $). $n$ is the number of rows in the matrix or the number of vertices in the graph. $m$ denotes the number of $NNZ$ (number of non-zero elements) in the matrix or the number of edges in the graph. The first array stores the number of $nnz$ for each row. These lengths are stored in a monotonically increasing array - the offset array. 
% The values between two consecutive entries in the array is equal to the $nnz$ for that row. 
The second array stores the actual $nnz$ elements. Through the offset array, it is possible to get the starting and ending point for the $nnz$ elements of a row.
Figure \ref{fig:graph-csr} illustrates a CSR representation of an example bipartite graph.
For discussing HashGraph, we will refer to the first array as $offset$ and the second array as $ keys $.

\paragraph{\bf Building HashGraph}
\label{sec:single_gpu_build}
At its core, building a HashGraph is just building both the $offset$ (hash values) and $keys$ arrays described above.
These are the only two arrays in a CSR representation, so these are the only two arrays in HashGraph.
The length of the $offset$ array is the number of possible hash values, i.e. $N/C$, while the length of $keys$ is the number of input keys to the hash table.
In this formulation, each index in $offset$ points to the first key in $keys$ that hashes to that index.
% \paragraph{Constructing $offset$ array for HashGraph}
% Note that $offset[i + 1] - offset[i] = deg(v_i)$.
% For HashGraph, $v_i$ corresponds to the number of input elements hashed to the hash value $i$. The construction of HashGraph is described in Alg. \ref{alg:sg-hg-build}, where the $offset$ array is constructed on line $1-10$. 
% This construction consists of the following two phases: 1) create an array with the degree of each hash and 2) prefix sum this array
% \paragraph{Constructing $keys$ array for HashGraph}
The last parallel \textit{for}-loop in Algorithm \ref{alg:sg-hg-build} illustrates describes how values are finally placed in the $keys$ array using the $offset$ array. 
% The $offset$ array gives the initial location in the hash-table for placing a specific hash value. 
The atomic instruction ensures that there is no collision for that specific entry.

\subsection{Efficient HashGraph with Bin Counting and its Application to Multi-GPU HashGraph }
In this subsection, we describe the second and more efficient HashGraph. Our multi-GPU HashGraph uses many concepts of this algorithm for preparing data for communication. Though we use them in a slightly different manner. 
% One of the most significant performance limiters for hash-tables is the large number of random memory accesses. 
The cost of random memory access is well-known for hash tables and has received significant attention. 
With Algorithm \ref{alg:sg-hg-build}, accessing an array at index $H_A[i]$ is effectively a random access, as $H_A[i]$ can vary widely depending on the hash function.

The efficient algorithm for HashGraph, first reorganizes $H_A$ so hash values generally lie near similar hash values, and then builds the hash table.
To do this in a cache-friendly manner, Green \cite{greenHashgraph2019} first assigns each hash value to one of $ B_L $ bins based on $H_A[i] / BinSize$ and then the hash values in order across all bins back into $H_A$.
Here, $B_L$ is small enough to fit in the cache.
All the phases of HashGraph use a single shared data-structure used by all threads. Using similar concepts, we extend this to multiple-GPUs.
% Thus, it can scale to thousands of threads. In the following subsections, we show how to extend this approach to a multi-GPU environment.

\begin{algorithm}[t]
    \tiny
\caption{Single-GPU HashGraph Build. Note $V = N / C$ \cite{greenHashgraph2019}}
\begin{algorithmic}[1]
\ParFor{$i = 0, \ldots, N - 1$}
    \State $H_A[i] = hash(A[i]) \mod V$
\EndParFor
\ParFor{$i = 0, \ldots, V - 1$}
    \State $CounterArray[i] = 0$
\EndParFor
\ParFor{$i = 0, \ldots, N - 1$}
    \State $AtomicAdd(CounterArray[H_A[i]], 1)$
\EndParFor
\State $offset = PrefixSum(CounterArray)$
\ParFor{$i = 0, \ldots, V - 1$}
    \State $CounterArray[i] = 0$
\EndParFor
\ParFor{$i = 0, \ldots, N - 1$}
    \State $pos = AtomicAdd(CounterArray[H_A[i]], 1)$
    \State $keys[pos + offset[H_A[i]]] = A[i]$
\EndParFor
\end{algorithmic}
\label{alg:sg-hg-build}
\end{algorithm}

\begin{algorithm}[h]
    \tiny
\caption{Multi-GPU HashGraph Build. Note $V = N_d / C$}
\begin{algorithmic}[1]
\ParFor{$d = 0, \ldots, DEVICES - 1$}
    \\ \Comment Phase 1 - Data Partitioning
    \ParFor{$i = 0, \ldots, N_d - 1$}
        \State $H_A[i] = hash(A[i]) \mod V$
    \EndParFor
    \ParFor{$i = 0, \ldots, N_d - 1$}
        \State $bin = H_A[i] / BinSize$
        \State $AtomicAdd(BinCounter[bin], 1)$ 
    \EndParFor
    \State $Reduce(BinCounter, 0...DEVICES - 1, D2H)$
    \State $BinOffset = PrefixSum(BinCounter)$
    \State $KeyCount = N_d / DEVICES$
    \State $HighestHash = KeyCount * (d + 1)$
    \State $HHIdx = BinSearch(BinOffset, HighestHash)$
    \State $BinSplits[d + 1] = HHIdx$
    \State $BCast(BinSplits, 0...DEVICES - 1, H2D)$
    \State $\_\_barrier\_\_$
    \\ \Comment Phase 2 - Data Reorganization
    \ParFor{$i = 0, \ldots, N_d - 1$}
        \State $min_d = Search(H_A[i], BinSplits)$
        \State $AtomicAdd(BuffCounter[min_d], 1)$ 
    \EndParFor
    \State $BuffOffset = PrefixSum(BuffCounter)$
    \ParFor{$i = 0, \ldots, DEVICES - 1$}
        \State $BuffCounter[i] = 0$
    \EndParFor
    \ParFor{$i = 0, \ldots, N_d - 1$}
        \State $min_d = Search(H_A[i], BinSplits)$
        \State $pos = AtomicAdd(BuffCounter[min_d], 1)$ 
        \State $Buffer[pos + BuffOffset[min_d]] = A[i]$
    \EndParFor
    \ParFor{$i = 0, \ldots, DEVICES - 1$}
        \State $FinalCounter[d] += BuffCounter[i]$
    \EndParFor
    \State $FinalOffset = PrefixSum(FinalCounter)$
    \State $\_\_barrier\_\_$
    % \ParFor{$i = 0, \ldots, DEVICES - 1$}
    %     \State $cudaMemcpy(FinalKeys + FinalOffset[i], Buffer_i + BuffCounter[d], D2D)$
    % \EndParFor
    \\ \Comment Phase 3 - Data Movement
    \State $AllToAll(FinalKeys, BuffCounter, D2D)$
    \\ \Comment Phase 4 - HashGraph Creation
    \State $BuildHashGraph(FinalKeys)$
\EndParFor
\end{algorithmic}
\label{alg:mg-hg-build}
\end{algorithm}

\subsection{Building Multi-GPU HashGraph}
At a high-level, multi-GPU  HashGraph has a HashGraph per GPU for a subset of the hash values. 
Each GPU has an associated \textit{hash range}, and that accelerator's HashGraph only stores keys with hashes within that range. 
Ideally, we want the number of keys on each device to be roughly equal for proper load balancing and storage utilization, i.e., $N / DEVICES$ keys.
This means we need to pick hash ranges per device such that roughly $N / DEVICES$ keys fall into each interval.
While it might seem like a daunting challenge to ensure that the ranges are near equal sizes, we benefit from the fact that hash functions do an excellent job of creating uniform distributions across a given range of integers.

%ODED - we need a plot here showing the examples discussed in the ``intuition text''
\subsubsection*{Multi-GPU Hash Table Intuition} - Recall that given a hash function $h$ and an input array $A$  of size $N$, for each entry $A[i]$, we typically compute the hash value to be $h(A[i])\%V$ (where $\%$ represents the {\it modulus} operation). Assuming our input has mostly unique elements, a good hash function will ensure that the number of elements hashed to a specific hash value will follow a uniform distribution. On average, each hash value will have $O(1)$ hash to it. 
For such an ideal distribution, we can divide the range $0...N-1$ into equal chunks and split them across the GPUs. Unfortunately, that is not the case as duplicates do exists in the input and hash-collisions can also occur. However, this is a good starting point and will work well for separate chaining
based hash tables.  
% In separate chaining, each hash-value is responsible for maintaining its list of input values \footnote{In open-addressing based hash tables, duplicates and collisions can overlap between multiple hash values. For the sake of simplicity, we primarily focus on separate chaining in our discussions.}. 

Instead of dividing the hash-range into chunks of equal size, consider splitting the hash-range so that each chunk will contain nearly the same number of entries (the range on each GPU varies). As it turns, this can be done using a prefix operation across the hash values followed by a binary search that will find the partition points. The challenge of computing a prefix operation in a distributed environment requires that each GPU maintain an array of length $N$ locally. This can prove to be prohibitive when $N$ is large and can easily exceed GPU memory. Therefore, we suggest using a binning strategy that effectively reduces the range and makes the prefix operation very viable. In the following subsections, we show that our new multi-GPU HashGraph uses two phases of binning. The first binning phase, which is at the entire system granularity, enables doing a prefix operation across the GPUs and enables good data partitioning. The second binning phase, which is done for each GPU and is a local computation, enables faster hash table creation using effective caching. %(similar to the approach by Green \cite{greenHashgraph2019}).

In the following subsections, we show that we can use a binning strategy to place input keys within a given hash-range into a single bin. 
% We then compute a distributed prefix operation efficiently on the bin sizes and show that we can split the bins in a near equal manner across the nodes.
Our first underlying assumption is that, if $BINS_G$ is the total number of bins, $BINS_G > DEVICES$. This is a fair assumption for most modern clusters with a few hundred to several thousand compute nodes. Our second assumption is that the average number of duplicate keys in our input is smaller than $\frac{N}{DEVICES}$. Without this assumption, we cannot ensure a perfect partitioning of the data. We do not take this assumption lightly. Most hash tables will face significant performance challenges when such a large number of duplicates exist in the input\footnote{In practice, if such a large number of duplicates exist in the input, then it is very likely that a hash table is note the ideal data-structure for fast lookups and scans.}. While a perfect-load balancing might not be achievable, HashGraph \cite{greenHashgraph2019} shows that it can deal with a large number of duplicates without a significant performance loss.

% ODED - Do you want to add "Phases" to your pseudo-code and then splut the discussion accordingly.
% ODED - I think that we want to distinguish between the bin counts for phase 1 (global) and phase 2 (local). 
%        What about using Bins_G and Bins_L to distinguish the two. Then in the experiment section we can state that we set Bins_G=Bins_L.
% ODED - do we want to use the term "rank" to refer to compute nodes? This better aligns with the standard terminology of the distributed community.

In the following, we explain the overall steps for the build process. These are associated with the pseudo-code in Algorithm \ref{alg:mg-hg-build}.
% The initial input is partitioned across the compute nodes.
% Note that the input is a set of keys on each device.
\paragraph{\bf Phase 1: Multi-GPU Data Partitioning} 

Our first goal is to partition the data across the multi-GPUs such that each GPU will be responsible for a near equal number of elements. To do this, we divide the hash range, $HR$, into $BINS_G$ sub-ranges. Each sub-range is a consecutive set of hash values and roughly covers $\frac{HR}{BINS_G}$ hash values. In most cases $HR= c \cdot N$ where $c \in [0.5,2]$.
$BINS_G$ is a tunable parameter. Choosing a small number of bins will lead to the bins having a large number of elements and make the partitioning harder (assuming a uniform distribution). Selecting a large number of bins will increase the memory footprint (as discussed above). We have found that setting $BINS_G~O(\sqrt{HR})$ does a good job in trading off the bin sizes, the number of elements in the bin, and the memory footprint.

After setting the value for $BINS_G$, each GPU creates an array of counters to count the number of instances per bin. In a first step, the counters' values get set to zero. Then each input within a given is hashed, and the counter for its respective bin is incremented. When this process has completed, each GPU has a distribution of values within the counters. This distribution is coarse grain and is for elements that fall into the same bin and not for the same hash-values. \emph{Also worth noting is that even if duplicates do exist, it is improbable that a single bin will be the target of all the duplicates. If it is the target of all the duplicates, then most traditional parallel and distributed hash tables will face the same type of challenges as our hash table; but ours is likely to perform better because of our collision management.}.  

% ODED to ALOK: Please verify that what I am writing below matches the pseudo code.
In the next phase, each GPU computes the prefix of the counters. An all-to-all (global) prefix sum operation across all GPUs follows the local prefix computation. With the global prefix sum array, we now have the distribution of values across the entire input. 
Recall that $BINS_G > DEVICES$. Our goal is to ensure that each GPU will receive roughly $\frac{N}{DEVICES}$ input keys.
Thus, each rank, $r$, will compute two binary searches in the global prefix sum array: $r \cdot \frac{N}{DEVICES}$ and $(r+1) \cdot \frac{N}{DEVICES}$. The entries found in the prefix sum array will enable defining the boundaries of the hash-ranges per rank.
% Algorithm \ref{alg:mg-hg-build} depicts this with the first two \texttt{for}-loops, which hashes all keys and bins them, respectively.

\paragraph{\bf Phase 2: Data Reorganization Per GPU}

With the sub-ranges each device will iterate over its data an additional time and count how many instances fall into each of the new sub-ranges (the number of sub-ranges is equal to the number of devices). 
Note that this phase also uses a prefix sum operation to create consecutive memory partitions for each device. These partitions will enable an efficient all-to-all communication phase where each device can share the relevant inputs with the remaining device---each device will send one message to all the other remaining devices.
The prefix operation in this phase is an array of length $DEVICES$, and its overhead is relatively small compared to other parts of the computation. 
Each device will now reorganize the input into $DEVICES$ partitions.
%\footnote{The reorganized data is also maintained in a CSR-like data structure.}. 
The values in the new partitions are not organized in any particular order. All we know about the values in the same partition is that these will be transferred to the same device for the hash graph creation.

% After computing the appropriate hash range per device, we must ship each key its corresponding device based on its bin.
% To do this, we construct a CSR per device.
% The $offset$ array is of length $DEVICES$, and the $keys$ array with the input keys on that device.
% This way, $offset[i]$ points to the keys meant for device $i$, allowing us to easily access device $i$'s keys.
% Algorithm \ref{alg:mg-hg-build} depicts this with third \texttt{for}-loop and the following prefix sum.

\paragraph{\bf Phase 3: Distributing keys to correct devices}
In this phase, the keys are sent to the device responsible for their specific hash-range.
This is equivalent to an all-to-all communication step and can is easily done because we stored keys per device with CSR.
When this has completed, each device will have all the input values that fall into the hash range under its responsibility. 
% CSR lets us access the exact keys to copy to destination device $i$ in $O(1)$ time and with a compact representation.
% Algorithm \ref{alg:mg-hg-build} illustrates with the final two \texttt{for}-loops.

% ODED - something is missing here in this subsection. This HashGraph is different than the single node HashGraph (the ranges are different)
% Can you add  (even in bullet points) what the exact differences are.
\paragraph{\bf Phase 4: Instantiating HashGraph per device}
After the all-to-all communication, each device has the correct set of keys per the device's hash range.
We can now instantiate a single-GPU HashGraph with these keys.
Together, each HashGraph across all devices forms the distributed HashGraph.

\paragraph{\bf Querying Single-GPU HashGraph}

HashGraph can be queried in the same manner as any other hash table.
The uniqueness of the HashGraph data structure has brought with it a new querying process. 
The query-set is the second set of values such that for each entry in the query set, we want to know if that value exists in the first set. A typical assumption is that the first set has been placed in a hash table (if not, a hash table for the first set is constructed). Now, each value in the query set gets hashed, and the hash table gets scanned for that value.

%First we have the set of keys that belong in the hash table, and second we have a set of \textit{query} keys we wish to lookup.
%Querying a multi-GPU HashGraph naturally extends querying a single-GPU HashGraph as described in Green \cite{greenHashgraph2019}.
We will first describe single-GPU HashGraph briefly, and then discuss how to extend this algorithm to the multi-GPU setting.

\label{sec:snsg-hg-query}
For a single-GPU HashGraph, querying the hash-graph is a simple two step process for querying.
Build a HashGraph for the query set. 
%We first build a HashGraph from the input keys, but we also build a second HashGraph for the set of query keys.
Given two HashGraphs, we can now perform list intersections between the corresponding lists of the two hash tables $HG_{1,i}$ with $HG_{2,i}$.
Due to the simplicity of HashGraph, we know exactly which keys hash to a particular value $i$ by iterating from $keys[offset[i]]$ to $keys[offset[i + 1]]$. This is especially efficient as the number of random memory accesses is reduced, the is cache re-use, and only the relevant elements are scanned. Green  \cite{greenHashgraph2019} shows that this is especially efficient when there the average number of elements per list is greater than 4 (a reasonable number). HashGraph continues to outperform existing hash tables as the number of duplicates grows. 
Lastly, note that each set intersection is independent of other intersections as there is no overlap between the lists, so running all the intersections is embarrassingly parallel and straightforward.

%Because of this, to lookup just the query keys that hash to $i$, we can run a set intersection on $keys[offset[i]\ldots offset[i + 1]]$ between each HashGraph's $keys$ and $offset$ arrays.
%By doing this for each hash value $i$, we successfully lookup each query key.
%In addition, for a good hash function, $offset[i + 1] - offset[i]$ tends to be small (on average $C$ values per $i$).
% Thus, running these set intersections can be done with massive parallelism on a GPU.

\paragraph{\bf Querying Multi-GPU HashGraph}
Querying multi-GPU HashGraphs is a natural extension of querying single-GPU HashGraphs. Similar to the single GPU algorithm, our multi-GPU querying algorithm consists of two steps: 1) build a multi-GPU HashGraph from the query keys and 3) run the intersections across all the GPUs.

It is crucial to note that both HashGraphs, input and query set, share the same hash range on each device.
After building these HashGraphs, we simply run set intersections on each single-GPU HashGraph in the same way described in \ref{sec:snsg-hg-query}
% Querying a Distributed HashGraph (DHG) is fundamentally the same as querying a single-GPU HashGraph.
% We build one DHG $DHG_A$ with the keys we want to store, and another DHG $DHG_B$ with the keys we wish to query.
% Then, just like in the single-GPU scenario, we intersect the list of keys associated with each hash value.
% Note that, when constructing $DHG_B$, we must use the same hash ranges per device used when building $DHG_A$.
% Doing this ensures that when we intersect a list from $DHG_A$ and a list from $DHG_B$, both lists will lie on the same device.

\subsection{Complexity Analysis}
\paragraph{\bf Work Complexity}
For this subsection, let $ N $ be the average number of keys per device, and $ P $ be the total number of devices.
Note that the total number of processed keys is thus $NP$.
The bottleneck of Algorithm \ref{alg:mg-hg-build} are lines $19-21$ and $27-30$, as these are \texttt{for}-loops that call a $Search$ function.
In our implementation, $Search$ is a linear search across all processes and has $O(P)$ work.
Since the outermost \texttt{for}-loop has $O(P)$ work, the inner \texttt{for}-loop has $O(N)$ work, and the $Search$ function has $O(P)$ work, the overall work complexity is $O(NP^2)$.
On most systems, however, $P$ tends to be very small compared to $N$.
For example, the NVIDIA DGX2 that we conduct our experiments, $P = 16$.
Since $P << N$, building a multi-GPU HashGraph is approximately $O(N)$ in practice.
\paragraph{\bf Space Complexity}
Each device stores a single HashGraph, i.e. a single CSR data structure.
While building a multi-GPU HashGraph involves instantiating other structures, such as $BinCounter$, $BinSplits$, and $FinalCounter$, these are lower-order terms in terms of memory.
Recall that a CSR structure is ultimately two arrays: one of length $N / C $ and one of length $ N $.
In total, this means that the space complexity of building a multi-GPU HashGraph is $O(N + N/C)$.
Since $C$ is not necessarily greater than $1$, this complexity cannot be reduced any further, as $N/C$ could be the dominant factor.

\setlength{\textfloatsep}{0pt}

\section{Experimental Setup}
\label{sec:experimental-setup.tex}

% \begin{table*}[t]
% \vspace*{-0.1 cm}
% \caption{GPU and CPU systems used in experiments.}
% \tiny
% \centering
% 
% \begin{tabular}{|c|c|c|c|c|c|c|c|c|c|c|c|c|c|} \hline
% System & $\#$ GPUS & Processor & Micro-architecture & SM & SP (per SM) & Total SPs & DRAM Size & DRAM Type &BW (GB/s)  & Form Factor & Power & Interconnect. & Interconnect BW (GB/s) \\ \hline \hline
% % DGX1     & 8 & V100  & Volta     & 80    & 64 & 5120 & 32GB    & HBM2 & 900 & SXM2 & 300 & NVLink & 160 \\  \hline
% DGX2     & 16& V100  & Volta     & 80    & 64 & 5120 & 32GB    & HBM2 & 900 & SXM3 & 350 & NV Switch & 300\\  \hline
% \end{tabular}
% 
% \vspace*{0.4 cm}
% 
% \begin{tabular}{|c|c|c|c|c|c|c|c|c|} \hline
% System & Architecture & Micro-architecture & Processor & Frequency & Cores & LL-Cache & DRAM Size & DRAM Type    \\ \hline \hline
% % DGX1 & CPU x86-64       & Broadwell   & $2 \times $ Intel Xeon E5-2698v4 & 2.1 GHz & $2 \times $ 20  & $2 \times $ 50 MB & 512GB  & DDR4 \\  \hline
% DGX2 & CPU x86-64       & Skylake   & $2 \times $ Intel Xeon Platinum 8168 & 2.7 GHz & $2 \times $ 24  & $2 \times $ 50 MB & 1536GB  & DDR4 \\  \hline
% \end{tabular}
% 
% \label{tab:gpu-cpu-systems}
% \end{table*}

\subsection{Systems and Configuration}
In this subsection, we discuss the different systems we use to evaluate our implementation.
We target NVIDIA V100 (Volta) GPUs for our implementation.
We also focus on two general systems: The NVIDIA DGX2 and the IBM POWER9 AC922 system.

\paragraph*{\bf{Volta GPUs}}

The V100 is a Volta (micro-architecture) based GPU with 80 SMs and 64 SPs per SM, for a total of 5120 SPs (lightweight hardware threads). In practice, roughly 40K software threads are necessary for fully utilizing the GPU.
The V100 has a total of 32GB of HBM2 memory and 6MB of shared-cache between the SMs. Each SM also has a configurable shared memory of $96KB$. 
%The V100 also has 640 tensor cores, though these are not used in our BFS algorithms. 
The V100 has two form factors: PCI-E and SXM. PCI-E is the de facto form factor found in most consumer GPUs. The SXM form factor is equivalent to placing a GPU on a board. The SXM form factor has multiple NVLink channels allowing GPUs to communicate with other multiple GPUs concurrently at a higher than PCI-E. The SXM form factor GPU is known for outperforming its PCI-E counterpart due to increased frequency and power consumption.
The V100 in NVIDIA's DGX-2 and IBM AC922 are slightly different. 
In the DGX-2 the NVIDIA V100 are the SXM3 form factors and have a peak power consumption of 350 watts. 
In the IBM AC922 the NVIDIA V100 are the SXM2 form factors and have a peak power consumption of 300 watts. 
%Details of the GPUs can be found in \ref{tab:gpu-cpu-systems}.

\paragraph*{\bf{DGX-2}}
The NVIDIA DGX2 server is a single node server with sixteen V100 GPUs. The DGX2 server was the server to introduce NVSwitch. NVSwitch enables communication from each to GPU to all the remaining GPUs for a total of 300GB/s of bandwidth. Each GPU has six incoming and outgoing links at 25GB/s (each). Thus, a GPU can send and receive 150GB/s concurrently. The fully connected network also ensures that the latency between the varying communication paths is uniform in length. The true benefit of NVSwitch over existing interconnects is the fact the on-device bandwidth and off-device bandwidth are within an order of magnitude of each other. This ensures that the communication within a DGX-2 is fairly balanced.
Ang \emph{et al.} \cite{li2019evaluating} give a detailed performance analysis of NVSwitch and NVLink.
The DGX-2 also has a high-end CPU processor, an Intel Xeon Platinum 8168 processor with 48 cores and 96 threads. The DGX-2 used in our experiments has 1.5TB of DRAM memory.
%  Additional details of the CPU can be found in \ref{tab:gpu-cpu-systems}.

\paragraph*{\bf{IBM POWER9 AC922}}
The IBM AC922 is a single node server with two POWER9 22C CPUs and six V100 GPUs.
Each of the V100s on the AC922 has $16$GB of memory, in contrast to the GPUs on the DGX2 which have $32$GB of memory.
These devices are divided equally between two sockets for one CPU and three GPUs per socket.
Within each socket, the three GPUs are connected to each other and to the CPU via NVLink 2.0, which has $100$GB/s total bidirectional bandwidth between two devices.
Between sockets, the AC922 uses IBM's X-bus as its interconnect, with a $64$GB/s bandwidth.
The AC922 also uses two POWER9 processors (one per socket) for a $36$ cores with $4$-way SMT.
The node also has $96$GB of HBM2 memory and $512$GB of DDR4 RAM.
% Further details can be found in \ref{tab:gpu-cpu-systems}.

\subsection{Input Sizes and Key Distribution}
We focus on $32$-bit keys in our implementation.
% These keys are integers sampled uniformly at random from $[1, N]$.
% Our input keys, consistent with single-GPU HashGraph, are sequential for each experiment.
% That is, our input keys are exactly $\{1,\ldots,2^k\}$.
Our input keys for each experiment fall in the range $\{1,\ldots,2^k\}$.
The precise value of $k$ depends on the experiment and is discussed in detail in Sec. \ref{sec:exp}
% We also use both the Murmur \cite{appleby2008murmurhash} hash function and the identical (ID) (i.e. $\text{hash}(x) = x$) hash function in our experiments.
In all of our experiments, we use the Murmur \cite{appleby2008murmurhash} hash function.
Both single-GPU HashGraph and WarpDrive use Murmur hash.

\subsection{Experiments Performed}
\label{sec:exp}
In this section, we outline each of the different experiments we run along with the frameworks we compare against.
For each experiment, we evaluate our implementation based on throughput, i.e. keys/sec.
This metric is the standard for evaluation in previous work \cite{greenHashgraph2019, junger2018warpdrive, barthels2017distributed}.
Note that for almost all of our experiments, we choose $C = 1$.
That is, our table size is equal to the number of input keys.
This is only false when we have over $2^{32}$ input keys, as our table size is always $32$-bit.
In these cases, we have a table size of $2^{31}$.
\paragraph*{\bf{Weak Scaling}}
% In our strong scaling experiments, we fix $N$ and observe how our throughput changes with device count.
% We run these experiments with various values of $N$, ranging from $2^{26}$ to $2^{33}$, depending on the experiment and memory capacity of the system.
We run weak scaling experiments to evaluate scalability, where we fix $N_d$ and observe how throughput changes with device count.
The value of $N_d$ depends on the system.
The V100s on the DGX2 have a greater memory capacity than the V100s on the AC922, allowing for larger experiments.
Our strong scaling results showed similar performance to our weak scaling results, but are omitted for brevity.

% \paragraph*{\bf{Index Tracking vs. Non-Index Tracking}}
% In some applications, for querying, it is sufficient only to know the existence of the query keys in the hash table. In other applications, the user needs to the index for the query key in the original input array $A$.
% This situation can arise if there is associated metadata at each index in $A$.
% We evaluate our implementation in both scenarios.
% In some experiments, we do not keep track of the index associated with each input key.
% We refer to these experiments as \textbf{Non-Index Tracking} runs.
% In others, we store and report the associated index for each input key if it is queried.
% We refer to these experiments as \textbf{Index Tracking} runs.
\paragraph*{\bf{Random vs. Sequential Keys}}
Consistent with single-GPU HashGraph, we run experiments where our input keys are sequential.
That is, for these experiments, we build a hash table with the keys $\{1,\ldots,2^k\}$ for some $k$.
We also evaluate mulit-GPU HashGraph on randomly generated key sets.
In these experiments, we randomly sample with replacement $2^k$ keys from the set $\{1,\ldots,2^k\}$.

% \paragraph*{\bf{ID vs. Murmur hashing}}
% As discussed above, we evaluate multi-GPU HashGraph on two hash functions: the identical hash function (ID), and the Murmur hash function \cite{appleby2008murmurhash}. The goal of using ID hash is that it places the inputs on the output device --- thus, the communication phase is trivial with no data moving across the network. Also, the binning phase on each of the GPUs is also trivial. Altogether, ID removes random memory accesses and communication making it typically faster.

% \paragraph*{\bf{Managed memory vs. Non-managed memory}}
% In our implementation, we must allocate space on each device for its single-gpu Hashgraph instance.
% We can naively run a \texttt{cudaMalloc()} during execution to handle this allocation.
% However, note that we know the total memory used across all devices beforehand as it is related to the total number of keys $N$.
% Thus, we can alternatively allocate the total space needed with \texttt{cudaMallocManaged()} in preprocessing and have each device fetch the exact amount of memory it needs.
% This solution is possible because of CUDA's Unified Memory.
% We show experiments with both allocation methods to illustrate the performance gain with the latter approach.

\paragraph*{\bf{Build vs. Querying}}
We show scaling experiments for both building a multi-GPU HashGraph given an input array $A$ along with querying a hash table with a query array.
In each of our experiments, the number of query keys equals the number of inputs keys.
Recall that querying a multi-GPU HashGraph, once the HashGraph of input keys is constructed, is two phases.
First, we build a multi-GPU HashGraph from the query keys, followed by a series of list intersections.

\paragraph*{\bf{Duplicate Keys}}
We also evaluate multi-GPU HashGraph with duplicate keys.
To that end, we vary the global hash range for our Murmur hash function while fixing the overall key count and GPU count.
We adjust the hash range simply by adjusting $V$, the value used in the modulus on line $4$ in Alg. \ref{alg:mg-hg-build}.
When adjusting our hash range, the average number of occurrences for any key is simply the ratio of key count of length of the hash range.

%\paragraph*{\bf{DGX2 vs. AC922}}
%As described above, we run all of our experiments on both the NVIDIA DGX2 and the IBM POWER9 AC922.

\setlength{\textfloatsep}{0pt}

\section{Performance Analysis}
\label{sec:results}
\begin{figure*}[ht!]
    \centering
    % \qquad
    \subfloat[DGX2 Build]{\includegraphics[scale=0.25]{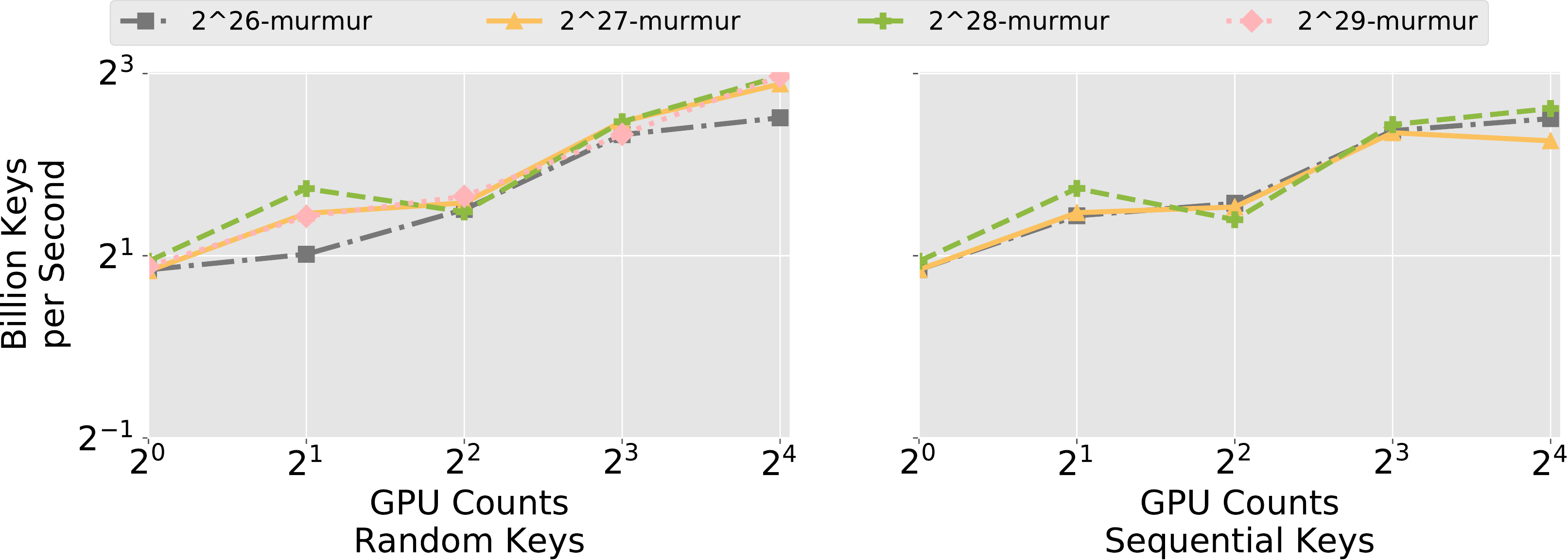}}
    % \qquad
    \hspace{1cm}
    \subfloat[DGX2 Query]{\includegraphics[scale=0.25]{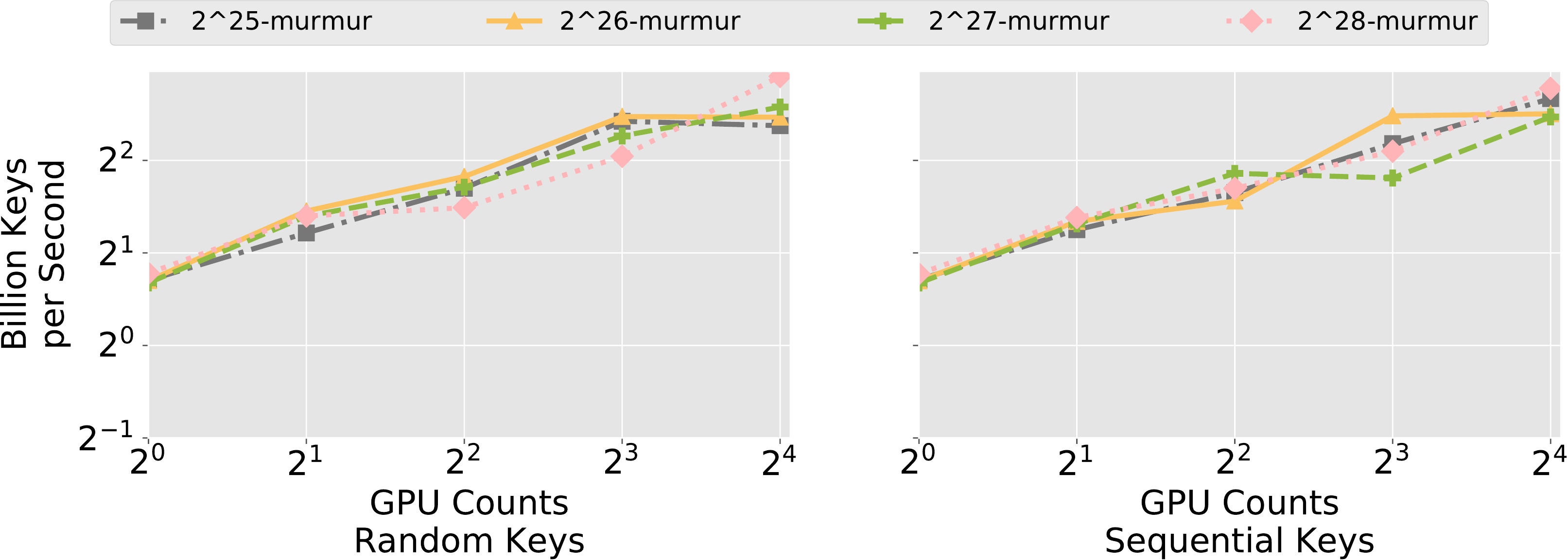}}
    % \qquad

    \subfloat[IBM Build]{\includegraphics[scale=0.25]{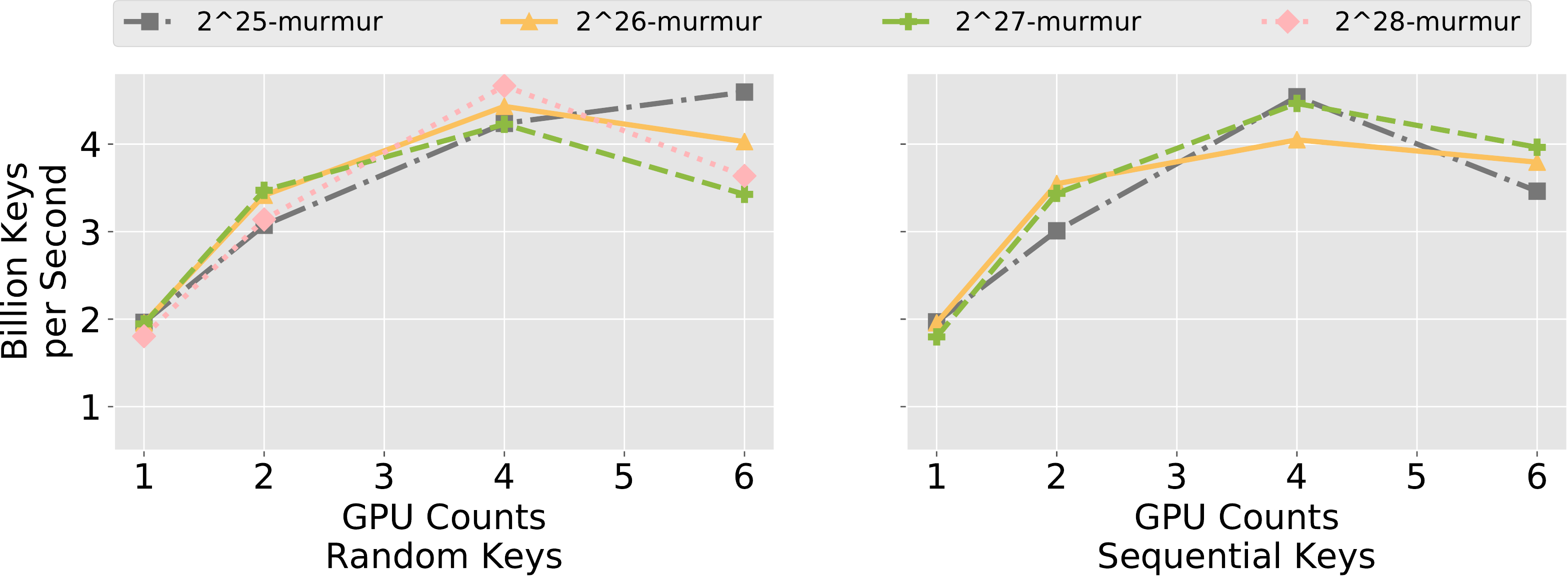}}
    % \qquad
    \hspace{1cm}
    \subfloat[IBM Query]{\includegraphics[scale=0.25]{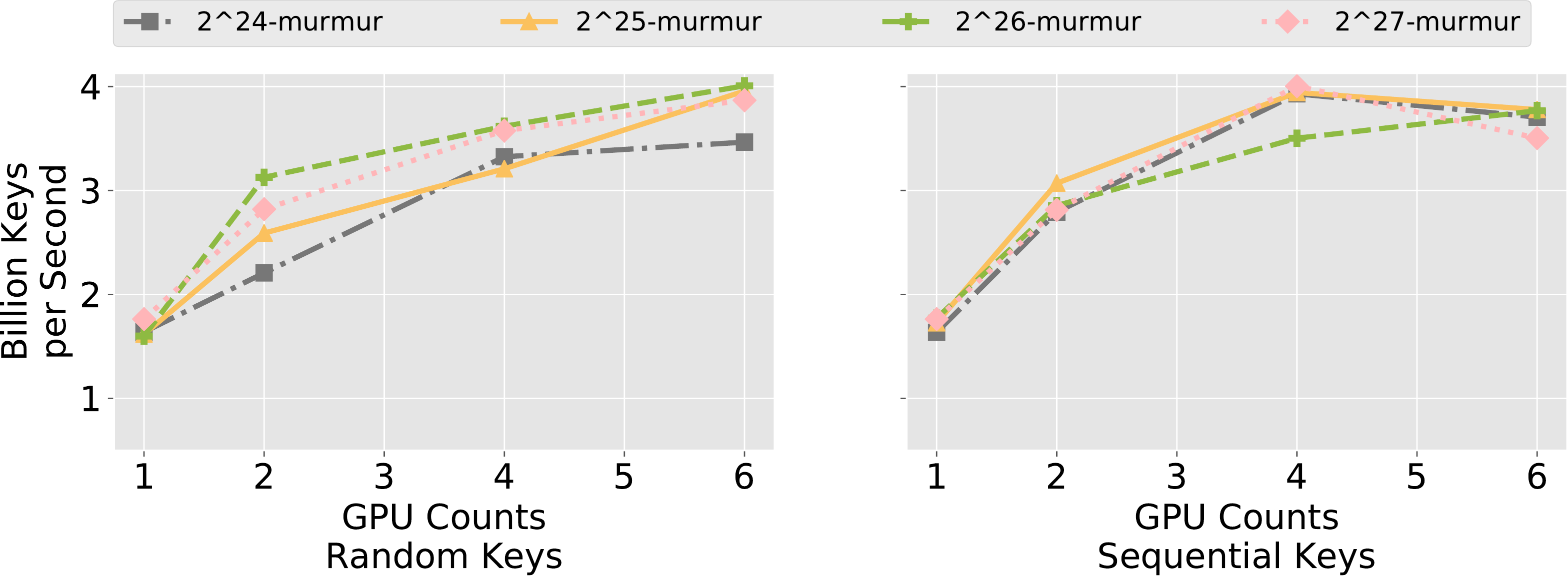}}
    % \qquad
    \caption{Weak scaling performance results on both DGX2 and IBM systems across GPU counts. $2^k$-\textbf{hash} refers to run with $2^k$ keys per device.}
    \label{fig:weak-scaling}
\end{figure*}

% \begin{figure*}[ht!]
%     \centering
%     % \qquad
%     \subfloat[DGX2 Build]{\includegraphics[scale=0.25]{figures/dgx2_strong_scaling_build-crop.pdf}}
%     \hspace{1cm}
%     % \qquad
%     \subfloat[DGX2 Query]{\includegraphics[scale=0.25]{figures/dgx2_strong_scaling_intersect-crop.pdf}}
%     % \qquad

%     \subfloat[IBM Build]{\includegraphics[scale=0.25]{figures/summit_strong_scaling_build-crop.pdf}}
%     % \qquad
%     \hspace{1cm}
%     \subfloat[IBM Query]{\includegraphics[scale=0.25]{figures/summit_strong_scaling_intersect-crop.pdf}}
%     % \qquad
%     \label{fig:strong-scaling}
%     \caption{Strong scaling performance results on both DGX2 and IBM systems across GPU counts. $2^k$-\textbf{hash} refers to run with $2^k$ keys and hash function \textbf{hash}. Datapoints not shown for a particular GPU count are because of out-of-memory failures.}
% \end{figure*}

\begin{figure*}[ht!]
    \centering
    % \qquad
    \subfloat[DGX2 - Build (left) and Query (right).]{\includegraphics[scale=0.25]{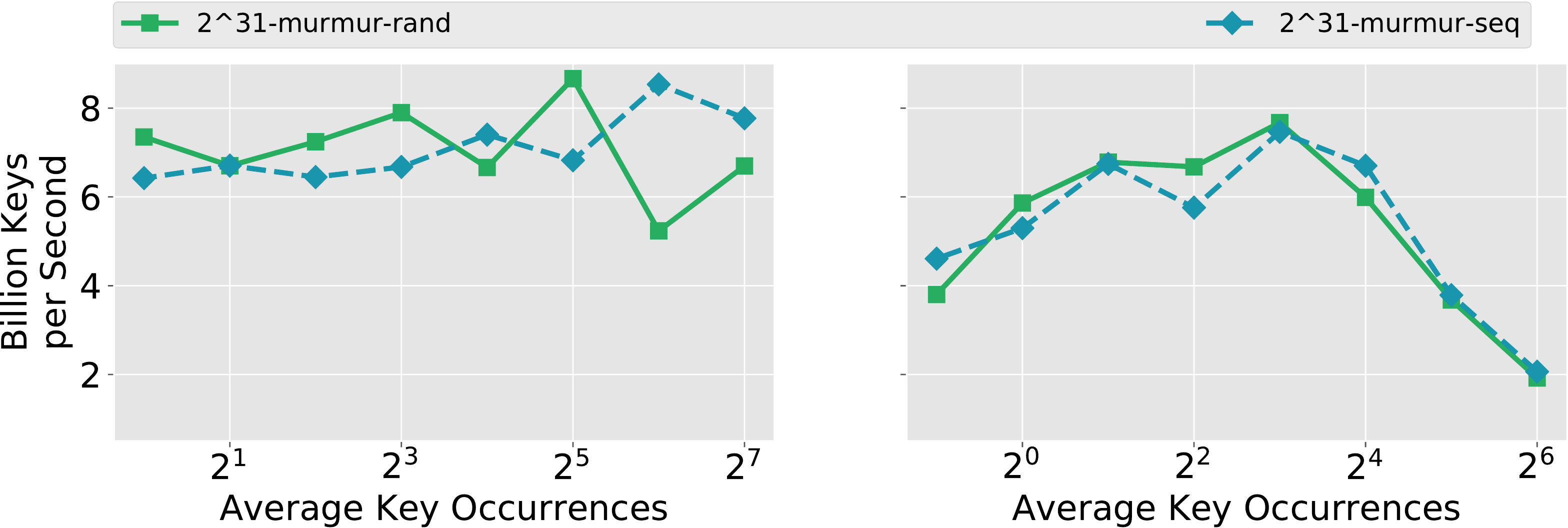}}
    % \qquad
    \hspace{1cm}
    \subfloat[IBM - Build (left) and Query (right).]{\includegraphics[scale=0.25]{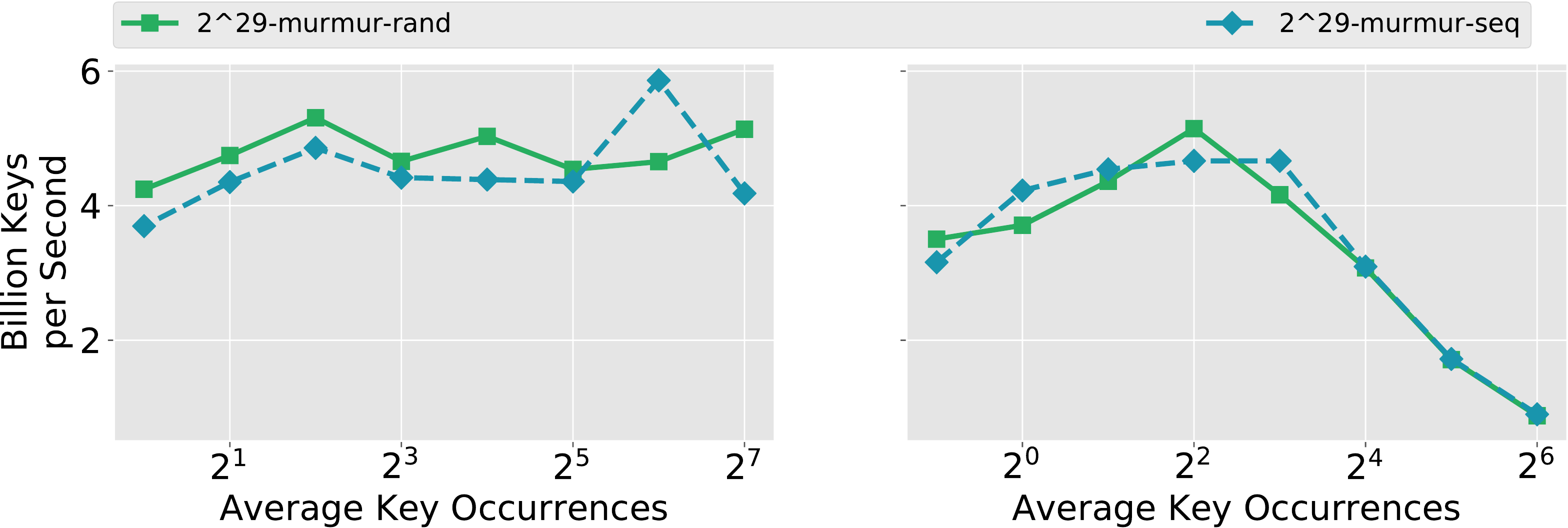}}
    \caption{Multi-GPU HashGraph performance on both DGX2 and IBM systems across average key occurrences. Average key occurrence is computed by dividing keycount by talbe size. $2^k$-\textbf{hash} refers to run with $2^k$ keys per device and hash function \textbf{hash}.}
    \label{fig:duplicate-keys}
\end{figure*}

\begin{figure*}[ht!]
    \centering
    \qquad
    \subfloat[DGX2]{\includegraphics[scale=0.25]{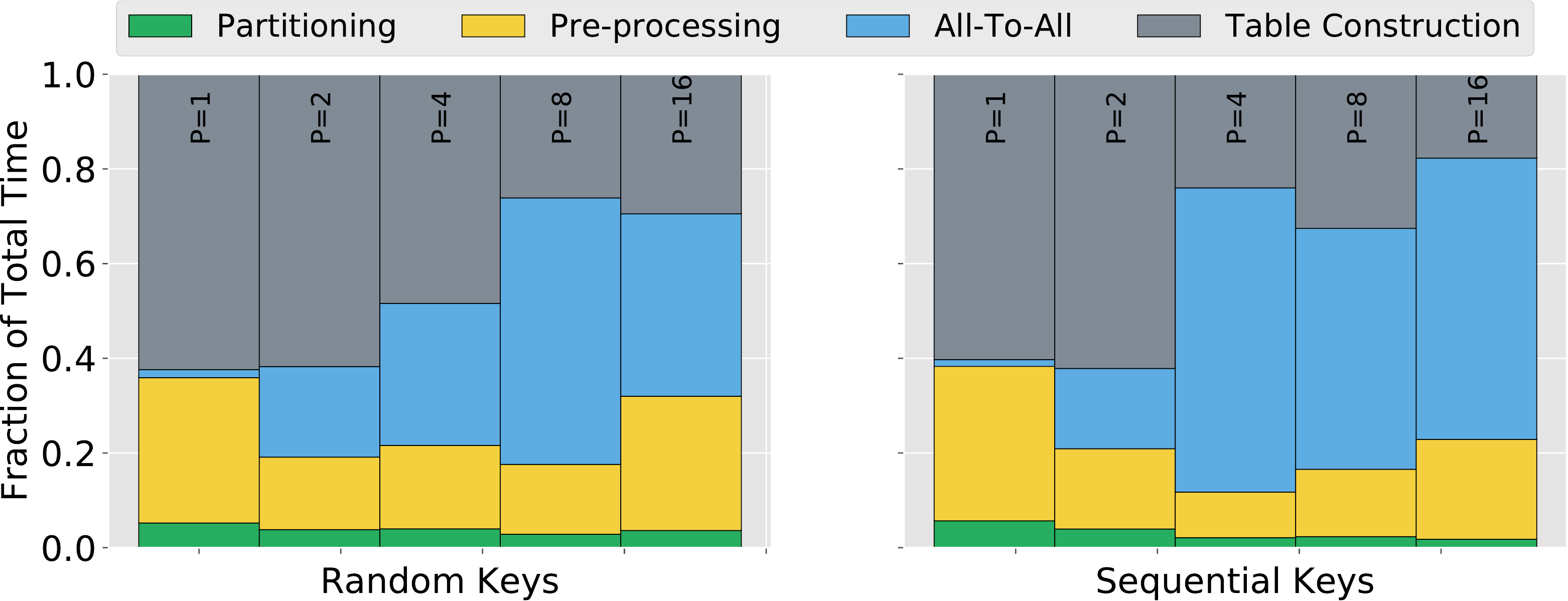}}
    \qquad
    \subfloat[IBM]{\includegraphics[scale=0.25]{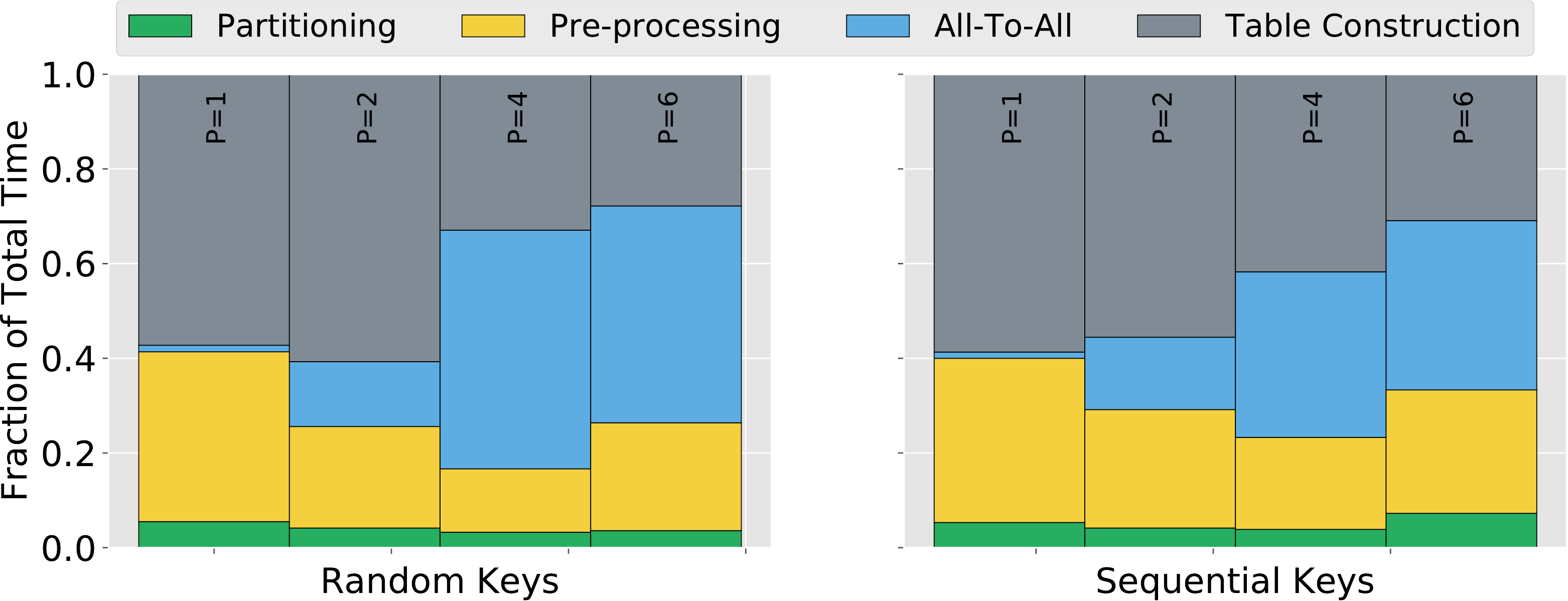}}
    \qquad
    % \caption{Weak scaling build performance breakdown on both DGX2 and IBM across GPU counts, for $2^{29}$ keys per GPU on DGX2 and $2^{28}$ keys per GPU on IBM. \textbf{Partitioning} refers to determining the hash range for each device (Phase 1 in Alg. \ref{alg:mg-hg-build}). \textbf{Pre-processing} refers to reorganizing keys on each device into CSR based on each key's destination GPU (Phase 2 in Alg. \ref{alg:mg-hg-build}). \textbf{All-to-All} refers to the all-to-all communication (Phase 3 in Alg. \ref{alg:mg-hg-build}), and \textbf{Table Construction} refers to building individual HashGraphs on each device (Phase 4 in Alg. \ref{alg:mg-hg-build}).}
    \caption{Normalized exeuction time for the four keys phases of our build algorithm. This is a weak scaling for $2^{29}$ and $2^{28}$ keys per GPU on DGX2 and IBM, respectively.}
    \label{fig:perf-breakdown}

\end{figure*}

\subsection{Weak Scaling}
The results of our weak scaling performance experiments are depicted in Fig. \ref{fig:weak-scaling}. 
The build operations shows good scalability on both the DGX-2 and the AC922. On the DGX-2 the build operation starts at about 1.5B \emph{keys/sec} and scales up to slightly over 6B \emph{keys/sec} and 8B \emph{keys/sec}, for 8 GPUs and 16 GPUs, respectively. That is an equivalent for a $4 \times$ and $6\times$ speedup over one GPU using our algorithm. The single GPU HashGraph \cite{greenHashgraph2019} is able to perform at roughly 2.3B \emph{keys/sec}. The reduction in performance of our new algorithm in comparison to the single-GPU algorithm is due to the second round of binning which requires random memory accesses. Altogether, the overhead of moving from a single-node algorithm to multi-node algorithm is not considerably high and accounts for about $35\%$. 

Lastly, our implementation uses CUDA's unified managed memory. We chose managed memory over unmanaged memory to avoid the cost of memory allocations during the execution of our algorithm. In our benchmarking, we saw allocating memory during execution led to inconsistent throughput. In contrast, managed memory gave fairly consistent performance at the cost of some performance where the GPU's runtime is responsible for fetching pages and moving them across devices. We include these runtime penalties in the times we report but note that these penalties account for some of the reduced speedups when going from 8 GPUs to 16 GPUs on the DGX-2. 

% \paragraph{ID vs. Murmur hash functions}
% Overall, like for strong scaling, multi-GPU HashGraph with ID hashing tends to be roughly $2\times$ faster than Murmur hashing.
% In Figure \ref{fig:perf-breakdown}, note that the all-to-all time is the largest difference between Murmur hashing and ID hashing.
% This is again due to the communication required by the all-to-all and the resulting page faults.

\paragraph{Phase Breakdown}

Fig. \ref{fig:perf-breakdown} depicts the breakdown of the four key phases four building our hash-table.
1) \textbf{Partitioning} determines the hash range for each device (Phase 1 in Alg. \ref{alg:mg-hg-build}. 
2) \textbf{Pre-processing} reorganizes keys on each device into a CSR based on each key's destination GPU (Phase 2 in Alg. \ref{alg:mg-hg-build}. 
3) \textbf{All-to-All} the all-to-all communication (Phase 3 in Alg. \ref{alg:mg-hg-build}.
4) \textbf{Table Construction} builds the individual HashGraphs on each device (Phase 4 in Alg. \ref{alg:mg-hg-build}.
This is a weak scaling for $2^{29}$ and $2^{28}$ keys per GPU on DGX2 IBM, respectively.

For both servers, the relative cost of the communication increases with the number of GPUs. However, the cause of the increase is for different reasons. As the all-to-all communication has $O(P^2)$ cost, going from 8 GPUs to 16 GPUs increases the number of messages by roughly $4\times$, while the average message decreases by $2\times$, the latency increases. NVIDIA's DGX-2 has NVSwitch which enables direct communication between all the GPUs. 
In contrast, the IBM AC922 requires that GPUs connected to the different CPU processors transfer date over NVLink and across the CPUs. This greatly reduces the all-to-all communication and explains why at 4 GPUs the relative cost of the communication for the AC922 is higher. The increase from 4 GPUs to 6 GPUs on the AC922 is not that high as the bottleneck does not change.
Lastly, we note that the cost of the preprocessing phase seems to grow with the number of GPUs. This seems to be an artifact of our implementation in how we place elements in bins. We plan on investigating other approaches for making this phase less expensive.

\paragraph{State-of-the-Art Comparison}
With our new algorithm, we see a build throughput of roughly $8B$ \emph{keys/sec} on DGX2 and $5B$ \emph{keys/sec} on the IBM AC922.
Note that Barthels et. al. \cite{barthels2017distributed} report having $10B$ keys / sec. when using $1024$ cores on their distributed inner-join algorithm.
They also report an $8B$ \emph{keys/sec}. throughput with $512$ cores, comparable to our throughput with Murmur hashing.
In place of using a distributed cluster, multi-GPU HashGraph is able to achieve comparable throughput on a single DGX2 server.
We can do so through a combination of 1) leveraging V100 accelerators on DGX2 with massive parallelism and 2) using communication primitives (e.g. all-to-all) that are typically prohibitively expensive for distributed settings, and 3) utilize NVIDIA's NVSwitch for fast data transfers.
Lastly, we note that we did not compare our algorithm to WarpDrive \cite{junger2018warpdrive} as its multi-GPU is not available. The single-GPU version of HashGraph outperforms WarpDrive.

\subsection{Strong Scaling}

Our strong scaling results exhibit the same performance trends as does our weak scaling, for the sake of brevity we do not show these duplicate plots. We do note that similar to weak scaling, we are able to test larger inputs on the NVIDIA DGX-2 due to the increased number of GPUs in comparison to the IBM AC922 and the difference in device memory between these two NVIDIA V100 GPUs.
% The throughput rates are nearly identical in these experiments.

\subsection{Build Vs. Query}
Overall the performance of our querying matches the performance of the single-GPU algorithm \cite{greenHashgraph2019}.
Recall that our query algorithm consists of two phases, building a second HashGraph and then doing a large number of set intersections. Thus, we expect our query algorithm to be slower than the building phase as it is one of two phases. As we need to store two HashGraphs, we could not build the same size hash tables and limited ourselves to hash tables one scale smaller than those used in the build experiments. 
Similar to the single-GPU algorithm, the intersect phase of the algorithm accounts for roughly $10\%$ of the execution time and the rest of the time is spent on building the hash-table for the query set. While the overhead of the building might seem high, there is also some motivation there. If one can speedup the performance of the build, the querying will also be faster. While this is obviously desirable, our algorithm on a DGX-2 already performs as well as 512 core CPU system.

% With two hash-tables being generated and the use of unified managed memory, we saw more page-faults than we saw when building a single table; which is why the performance is not identical to that in the single GPU version. 
% We have not been able to determine a good way to capture the cost of a page-fault.

\subsection{Duplicate Keys}
One of the nice features of HashGraph \cite{greenHashgraph2019} is its ability to deal with a large number of collisions created by duplicate keys within the input.
Fig. \ref{fig:duplicate-keys} depicts the performance of our new multi-GPU hash-table as a function of the number of times that each key appears in the input. Moving on the x-axis from left to right the number of times that the key appears in the input increases.
Note, the build rate is fairly consistent for all duplicate rates, as expected.
The query plot has a bit more variation. It starts off with a fairly consistent throughput, until the number of duplicates is roughly $2^3=8\times$. From this point onwards, the throughput decreases at a steady rate. This is not surprising as due to duplicates, many lists will have $2^i$ duplicates. 
As our intersection operation counts the number of time each key appears in the input, we are required to iterate through the entire list. Thus, for each entry in the query set we need to do $2^i$ comparisons. Since the query set also has duplicates and its lists are now longer, the total time per intersection increases in a quadratic manner. Which explains the fast decay in throughput. On a positive note, this matches the trend of the single GPU. The single GPU version was over $100\times$ than other state-of-the-art hash-tables for similar duplicate rates.

\setlength{\textfloatsep}{0pt}

\section{Conclusions}
In this paper, we presented a new algorithm for building a single-node hash table with performance comparable to distributed hash tables.
We accomplished this by extending an earlier single-GPU hash table, HashGraph, to the multi-GPU setting.\cite{greenHashgraph2019}.
We enhanced single-GPU HashGraph in three ways.
First, we extended a binning approach used in single-GPU HashGraph to effectively partition keys between GPUs.
Second, used additional sparse-graph data structures to organize keys for proper communication. 
Lastly, we leveraged communication primitives that are typically prohibitive in distributed settings.
We evaluate our algorithm's performance on two different systems: an NVIDIA DGX-2 server with 16 GPUs, and an IBM AC922 with 6 GPUS. 
Our algorithm shows promising scalability. 
In particular, we show that our algorithm can scale to hundreds of thousands of cores available on 16 GPUs.
We also showed that the various phases of our new algorithm are scalable and can be executed with hundreds of thousands of threads.
Overall, we present a multi-GPU hash table can process $8B$ keys per second, comparable to some CPU-based distributed hash tables with $500-1,000$ cores.
% In this paper we showed a new algorithm for building a distributed hash table which extends HashGraph from single-GPU to multi-GPU. 
% We show that we can use the binning technique first introduced for the single-GPU algorithm for implementing a simple and scalable communication mechanism for transferring the keys to their destination. 
% We use an equivalent of two rounds of binning, the first for preparing the keys for the communication phase and the second for building a subsection of the hash-table on each of the compute nodes.
% 
% 
% We show performance of our algorithm on two different systems with multiple GPUs: an NVIDIA DGX-2 server with 16 GPUS and an IBM AC922 with 6 GPUS. Our algorithm shows promising scalability. We showed that the various phases of our new algorithm are scalable and can be executed with hundreds of thousands of threads.
% 
% The performance of our hash-table on the DGX-2 is equal to the performance of a CPU-based cluster with $500-1,000$ cores.
% 

\setlength{\textfloatsep}{0pt}

% \affil[1]{Computational Science and Engineering, Georgia Institute of Technology - USA}

\thispagestyle{plain}
\pagestyle{plain}

\bibliographystyle{siam}
\bibliography{bibfile, hash}

\end{document}